\documentclass[sigconf,screen,authorversion]{acmart}

\usepackage{balance}       % to better equalize the last page
\usepackage{graphics}      % for EPS, load graphicx instead 
\usepackage{color}
\usepackage{soul}
\usepackage{xcolor}
\usepackage{booktabs}
\usepackage{textcomp}
\usepackage{multirow}
\usepackage{float}
\usepackage{listings} % for code snippets
\usepackage{amsmath,bm}
\usepackage{pifont}
\usepackage[labelformat=simple]{caption}
\usepackage{subcaption}
\usepackage[ruled,vlined,linesnumbered]{algorithm2e}
\usepackage{array}  

\usepackage{enumitem}
\usepackage{microtype} 
\usepackage{ccicons} 
\usepackage[disable]{todonotes}

\definecolor{IA}{RGB}{180,219,108}
\definecolor{SEL}{RGB}{161,226,49}
\definecolor{VT}{RGB}{180,219,108}
\definecolor{DP}{RGB}{222,213,65}
\definecolor{DF}{RGB}{215,207,108}
\definecolor{ID}{RGB}{229,229,42}
\definecolor{DT}{RGB}{192,213,56}
\definecolor{GM}{RGB}{32,216,253}
\definecolor{LM}{RGB}{243,166,72}
\definecolor{TR}{RGB}{172,184,241}

\newcommand{\ttt}[1]{{\tt{#1}}}

\newcommand{\tool}{\textsc{Ivie}}

\definecolor{javared}{rgb}{0.6,0,0} % for strings
\definecolor{javagreen}{rgb}{0.25,0.5,0.35} % comments
\definecolor{javapurple}{rgb}{0.5,0,0.35} % keywords
\definecolor{javadocblue}{rgb}{0.25,0.35,0.75} % javadoc
\definecolor{lightblue}{rgb}{0.63, 0.79, 0.95}
\definecolor{lightgray}{gray}{0.9}

\lstdefinestyle{MyJavaSmallStyle} {
  language=Java,
  frame=lines,
  xleftmargin=15pt, 
  stepnumber=1, 
  numbers=left, 
  numbersep=5pt,
  stepnumber=1,
  numberstyle=\tiny\bf,%\color[gray]{0.777}, 
  belowcaptionskip=\bigskipamount,
  captionpos=b, 
  escapeinside={*'}{'*},
  tabsize=5,
  emphstyle={\bf},
  basicstyle=\scriptsize\ttfamily,
  keywordstyle=\color{javapurple}\bfseries,
  stringstyle=\color{javared},
  commentstyle=\color{javagreen},
  morecomment=[s][\color{javadocblue}]{/**}{*/},
  showspaces=false,
  columns=flexible,
  showstringspaces=false,
  morecomment=[l]{//},
  tabsize=2,
  morekeywords={, Package,Invariant,Class,Method,Field,Where,Assert,ToLc,Split,Msg,Immutable,<<<,eq,neq,not,has,Assert,AssertExists,Attribute,Uc,Lc,},
  breaklines=true
}

\lstdefinestyle{MySimpleStyle} {
  xleftmargin=15pt, 
  stepnumber=1, 
  numbers=none, 
  numbersep=5pt,
  stepnumber=1,
  belowcaptionskip=\bigskipamount,
  captionpos=b, 
  escapeinside={*'}{'*},
  tabsize=5,
  emphstyle={\bf},
  basicstyle=\small\ttfamily\selectfont,
  keywordstyle=\color{javapurple}\bfseries,
  stringstyle=\color{javared},
  commentstyle=\color{javagreen},
  morecomment=[s][\color{javadocblue}]{/**}{*/},
  showspaces=false,
  columns=flexible,
  showstringspaces=false,
  morecomment=[l]{//},
  tabsize=2,
  morekeywords={, Package,Invariant,Class,Method,Field,Where,Assert,ToLc,Split,Msg,Immutable,<<<,eq,neq,not,has,Assert,AssertExists,Attribute,Uc,Lc,},
  breaklines=true
}

\newcommand{\rev}[1]{{#1}}

\definecolor{andrewpurple}{HTML}{A53DFF}
\definecolor{andreworange}{HTML}{E07400}
\definecolor{darkgreen}{HTML}{009B55}
\definecolor{darkblue}{HTML}{004d80}
\definecolor{magenta}{HTML}{99195d}

\AtBeginEnvironment{quote}{\itshape}

\def\subparagraph#1{\textbf{#1.}}

\def\shortspace{\kern 0.1em}

\def\KaTeX{K\kern-.2em\raisebox{.2em}{\scriptsize A}\kern-.12em\TeX}

\definecolor{niceblue}{HTML}{8295ff}
\def\bigbox{\color{niceblue}\rule[.25ex]{1ex}{1ex} \hskip .1ex}
\def\smallbox{\hskip .25ex \color{gray}\rule[.5ex]{.5ex}{.5ex} \hskip .25ex \hskip .1ex}
\def\boxes#1#2{
\hskip .1ex % Add a bit of space, because this seems necessary for the boxes not to take up the whole line?
\newcount\boxnum
\boxnum=0
\loop
\ifnum \boxnum<#1 \bigbox \else \smallbox \fi

\advance \boxnum by 1
\ifnum \boxnum<#2
\repeat
}

\newcolumntype{L}[1]{>{\raggedright\let\newline\\\arraybackslash\hspace{0pt}}m{#1}}

\newenvironment{inlinefigureenv}
{\setlength{\topsep}{2.5ex}\center}
{\endcenter}

\AtBeginDocument{%
  \providecommand\BibTeX{{%
    \normalfont B\kern-0.5em{\scshape i\kern-0.25em b}\kern-0.8em\TeX}}}

\setcopyright{acmcopyright}
\copyrightyear{2024}
\acmYear{2024}
\setcopyright{rightsretained}
\acmConference[CHI '24]{Proceedings of the CHI Conference on Human Factors in Computing Systems}{May 11--16, 2024}{Honolulu, HI, USA}
\acmBooktitle{Proceedings of the CHI Conference on Human Factors in Computing Systems (CHI '24), May 11--16, 2024, Honolulu, HI, USA}
\acmDOI{10.1145/3613904.3642239}
\acmISBN{979-8-4007-0330-0/24/05}

\begin{document}

\title{\tool{}: Lightweight Anchored Explanations of Just-Generated Code}

\settopmatter{authorsperrow=4}
\author{Litao Yan}
\email{ltyan@seas.upenn.edu}
\orcid{0009-0009-5077-354X}
\affiliation{%
  \institution{University of Pennsylvania}
  \city{Philadelphia}
  \state{PA}
  \country{USA}
  \postcode{19104}
}

\author{Alyssa Hwang}
\email{ahwang16@seas.upenn.edu}
\orcid{0009-0006-4827-8505}
\affiliation{%
  \institution{University of Pennsylvania}
  \city{Philadelphia}
  \state{PA}
  \country{USA}
  \postcode{19104}
}

\author{Zhiyuan Wu}
\email{wuzed@seas.upenn.edu}
\orcid{0009-0001-8016-5985}
\affiliation{%
  \institution{University of Pennsylvania}
  \city{Philadelphia}
  \state{PA}
  \country{USA}
  \postcode{19104}
}

\author{Andrew Head}
\email{head@seas.upenn.edu}
\orcid{0000-0002-1523-3347}
\affiliation{%
  \institution{University of Pennsylvania}
  \city{Philadelphia}
  \state{PA}
  \country{USA}
  \postcode{19104}
}

\renewcommand{\shortauthors}{Yan et al.}

\begin{abstract}
% Edited to remove dangling words
Programming assistants have reshaped the experience of programming into one where programmers spend less time writing and more time critically examining code. In this paper, we explore how programming assistants can be extended to accelerate the inspection of generated code. We introduce an extension to the programming assistant called \tool{}, or \emph{instantly visible in-situ explanations}. When using \tool{}, a programmer's generated code is instantly accompanied by explanations positioned just adjacent to the code. Our design was optimized for low-cost invocation and dismissal. Explanations are compact and informative. They describe meaningful expressions, from individual variables to entire blocks of code. We present an implementation of \tool{} that forks VS Code, applying a modern LLM for timely segmentation and explanation of generated code. In a lab study, we compared \tool{} to a contemporary baseline tool for code understanding. \tool{} improved understanding of generated code, and was received by programmers as a highly useful, low distraction complement to the programming assistant.

% Programming assistants have reshaped the experience of programming into one where programmers spend less time writing and more time critically examining code. In this paper, we explore how programming assistants can be extended to accelerate the inspection of generated code. We introduce an extension to the programming assistant called \tool{}, or \emph{instantly visible in-situ explanations}. When using \tool{}, a programmer's generated code is instantly accompanied by explanations positioned just adjacent to the code. Our design was optimized for extremely low-cost invocation and dismissal. Explanations are compact and informative. They describe meaningful expressions, from individual variables to entire blocks of code. We present an implementation of \tool{} that forks VS Code, applying a modern LLM for timely segmentation and explanation of generated code. In a lab study, we compared \tool{} to a contemporary baseline tool for code understanding. \tool{} improved understanding of generated code, and was received by programmers as a highly useful, low distraction, desirable complement to the programming assistant.
\end{abstract}

\begin{CCSXML}
<ccs2012>
   <concept>
       <concept_id>10003120.10003121.10003129</concept_id>
       <concept_desc>Human-centered computing~Interactive systems and tools</concept_desc>
       <concept_significance>500</concept_significance>
       </concept>
 </ccs2012>
\end{CCSXML}

\ccsdesc[500]{Human-centered computing~Interactive systems and \nolinebreak tools}

\keywords{Programming assistants, instructive copilots, anchored explanations, comprehension support, variable levels of detail, brevity, easy invocation, easy dismissal, label overlays}

\begin{teaserfigure}
  \includegraphics[width=\textwidth]{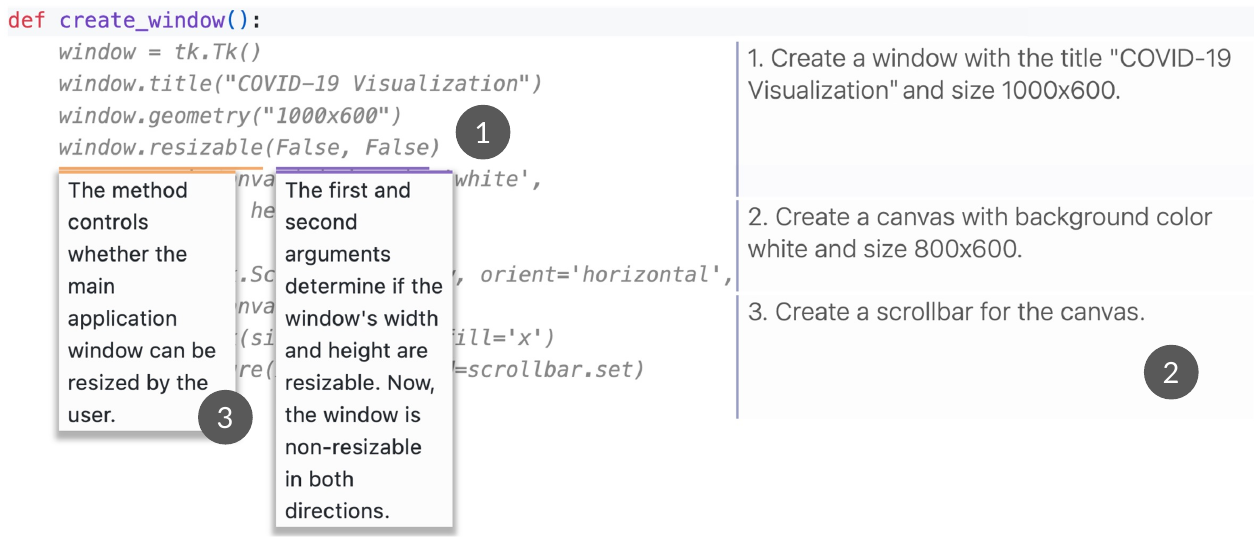}
  \vspace{-4ex}
  \caption{\tool{} augments the interactive programming assistant with instant explanations that help programmers examine generated code. \textmd{When a programming assistant suggests code (\textit{italic} text above, \textcolor[HTML]{595959}{\ding{202}}), \tool{} annotates it with brief, informative explanations. Explanations appear at the level of blocks of code (in the right margin, \textcolor[HTML]{595959}{\ding{203}}) and expressions (anchored beneath the line the programmer hovers over, \textcolor[HTML]{595959}{\ding{204}}). For single-line suggestions, expression explanations appear automatically. \tool{}'s explanations help programmers break up complex or unfamiliar suggestions into pieces that can be more readily understood.}}
  \Description{A code editor, containing a function named “create_window.” Below the function name, Copilot has generated a 15-line code suggestion. The suggestion is accompanied by automatically-generated explanations, which are shown in overlays next to the code. Explanations are shown for both multi-line segments of code, and for individual expressions (e.g., describing the function and and arguments involved in an API call that the user is hovering their mouse over).}
  \label{fig:teaser}
  \vspace{2ex}
\end{teaserfigure}

\maketitle

\section{Introduction}
\label{sec:introduction}

Since their recent release, programming assistants have begun to reshape the process of writing code. Programming assistants are a kind of interactive programming aid that help a programmer write code. The predominant interaction model is exemplified by GitHub Copilot~\cite{copilot}. Under this model, the programming assistant anticipates code that the programmer is going to write---either the remainder of a line, or even many lines at once---and proposes that code to the programmer. A programmer can then accept the code with a simple command, like pressing the ``Tab'' key. Production programming assistants have become very good at proposing code, even long chunks of code, should the programmer's code resemble something seen in the massive corpora of code that the programming assistant was trained on.

This interaction model has proven to be both desirable and useful. Programming assistants have seen massive adoption. Copilot estimates adoption by 1.2 million users~\cite{GitHubCo}. They have seen integration into some of the most commonly-used code editors such as VS Code. And they are used to generate a lot of code: recent estimates suggest that as much as 30\%~\cite{liang2023understanding} or 40\%~\cite{GitHubCo} of a programmer's code originates from programming assistants when in active use. The widespread and frequent usage of programming assistants suggests that they may be here to stay.

For the effort that they save, programming assistants introduce another kind of effort. Namely, programming assistants eliminate effort writing code and replace it with effort reviewing code~\cite{bird2022taking}. Studies have reported programmers spending an undesired amount of time understanding and debugging generated code~\cite{barke2023grounded,vaithilingam2022expectation}; they spend time examining generated code~\cite{barke2023grounded}, often examining details of generated code's logic in depth~\cite{liang2023understanding}. Obstacles to understanding generated code include its use of previously-unknown APIs or methods, structural complexity~\cite{liang2023understanding}, and the length of generated code~\cite{liang2023understanding,barke2023grounded}. For particularly long generations, programmers have described themselves as ``distracted by everything [the assistant] is throwing at [them],'' ``lost in the sauce,'' and ``discombobulated''~\cite{barke2023grounded}. Perhaps troublingly, programmers have reported understanding less of how and why generated code works~\cite{bird2022taking}, with one lab study suggesting that programmers apportion less attention to generated code than that written by human authors~\cite{al2022readable}.

This raises the question of whether our modern notion of the programming assistant is incomplete. A widely-used metaphor for  programming assistants is that they are copilots. If a programming assistant is a copilot, it is one that does its work without explaining its actions. Its actions are visible, but the copilot makes no effort to make them understandable. This is desirable when the copilot's actions are familiar, like when it generates code with known idioms. However, it is less desirable when the copilot's actions are unfamiliar, like when it generates code with unfamiliar APIs and structures. If programmers are expending effort to understand generated code, perhaps the notion of the copilot needs to be expanded to one of an \emph{instructive copilot} that prioritizes the programmer's understanding by explaining its work.

In this paper, we explore this notion of the instructive copilot. We develop a tool, \tool{}, which provides lightweight, anchored, AI-generated explanations of just-generated code. When using an editor with \tool{}, programmers see generated code instantly accompanied by brief labels that explain what the code does~(Figure~\ref{fig:teaser}). These labels are meant to answer questions like ``what does this parameter do?'' or ``what does this segment of generated code do?'' That is, they are tailored to helping programmers understand unfamiliar APIs and idioms in the generated code. We call the labels \emph{instantly visible in-situ explanations} (i.e., \tool{}). The labels are designed for tight integration into the modern workflow of generating, reviewing, and accepting code; they are meant to eliminate extraneous time it takes to provide answers to basic questions about the code prior to the programmer accepting or rejecting the code. The labels are very easy to invoke and dismiss. They explain code at the level of both blocks (for long, multi-line generations) and expressions (for dense one-liners). By providing instant, brief, anchored explanations, \tool{} seeks to remove information costs intrinsic to comparable forms of support for understanding generated code like chatbot-based programming helpers (e.g.,~\cite{copilotX, GenieAI, nam2023inide}) or code explainers that require code selection (e.g.,~\cite{EasyCode, GitHubNe24:online}). 

\tool{} has a straightforward architecture, leveraging an off-the-shelf LLM to segment code into explainable units (at both the expression and block level), and create brief explanations of those units. Most of \tool{}'s code consists of presentation logic that overlays LLM outputs as anchored explanations in the editor. We demonstrate the viability of bringing \tool{} to production editors by implementing it in the widely-used Visual Studio Code~\cite{VisualSt7:online}. 

We present a detailed in-lab usability study of \tool{}. The study explores \tool{}'s impact on the experience of understanding generated code containing unfamiliar APIs. The study compared \tool{} to a contemporary AI-based help-seeking baseline, namely a GPT-based in-editor chatbot that can answer questions about selected code. Our study shows that there is a clear place in the ecosystem of programming tools for \tool{}: it was preferred to the chatbot by nearly all participants. \tool{} led to improved comprehension of generated code, while decreasing perceived task load. Participants described \tool{} as being highly useful without being distracting. They characterized \tool{} as a complementary addition to a modern programming workspace alongside documentation and chatbot help.

In summary, this paper contributes:
\begin{itemize}
\item The notion of an \emph{instructive copilot} as a programming assistant that not only generates code, but also supports its understanding with timely, anchored, brief explanations.
\item The implementation of this idea in {\tool}, an extension to the programming assistant that augments generated code with overlay descriptions that are brief, informative, instantly visible, and easy to dismiss.
\item A usability study that confirms \tool{}'s value as an desirable complement to a programming assistant that improves code comprehension with low task load and distraction.
\end{itemize}
\section{Background and Related Work}

\subsection{User experience of programming assistants}
In recent years, a number of tools have incorporated large language models (LLMs) to provide production-level assistance for generating and describing code. These tools, often called ``programming assistants,'' include GitHub Copilot~\citep{copilot}, OpenAI GPT-4~\citep{openai2023gpt4}, Amazon CodeWhisperer~\citep{AICodeGe98:online}, IntelliCode Compose~\citep{svyatkovskiy2020intellicode}, and CodeT5+~\citep{wang2023codet5}. As programming assistants have been adopted by a growing number of programmers, researchers have begun to examine their effect on  programming activity. Some of these studies have characterized the tools' influence on how programmers understand code, particularly generated code~\citep{al2022readable, vaithilingam2022expectation, barke2023grounded, mozannar2023reading, xu2021inide, liang2023understanding}. We detailed takeaways from these studies in detail in Section~\ref{sec:introduction}; in brief, programmers report understanding less of the generated code than they do of the code they write, spending time examining the code, and sometimes finding the generated code difficult to read. These observations motivate our focus on developing aids to support the reading of generated code.

A recent study by \citet{barke2023grounded} characterized interaction with programming assistants as consisting of two modes: acceleration, where the assistant helps a programmer write code of the kind they already have some idea how to write; and exploration, where the programming assistant assists a programmer in determining and carrying out goals. We posit that a tool like \tool{} could be being useful in supporting both modes of interaction. As our results suggest in Section ~\ref{sub:understand}, explanations of the kind \tool{} provides can be helpful in learning about the behavior of unfamiliar code that might be generated during exploration, as well as in providing a timely comprehension aid for longer and more complex code that might appear in the middle of acceleration.

\subsection{Program comprehension}
\label{sec:comprehension_related}
Program comprehension---or the process of understanding computer code---is an essential and frequent programming activity. In one study, program comprehension accounted for as much as 50\% of programmers' time~\cite{xia2017measuring}. Program comprehension has been researched in depth by the HCI and software engineering research communities (see for instance reviews by \citet{detienne2001software} and \citet{crichton2022revisiting}). In some theories of program comprehension (e.g.,~\cite{detienne2001software}), comprehension relies on the identification of schemas---or meaningful structures---in the code, and synthesis of those schemas into a mental model of the program. We see the role of \tool{} as recognizing unfamiliar schemas in code on a programmer's behalf, and explaining them in approachable terms.

Program comprehension is sometimes considered as taking place either top-down or bottom-up. When following a top-down approach, programmers form hypotheses about a program's intent, and then verify those hypotheses by looking for recognizable features or ``beacons'' in the code~\citep{brooks1983towards,5010283}. When following a bottom-up approach, programmers progressively group units of the code into larger and larger abstractions~\citep{von1993program,pennington1987stimulus,shneiderman1979syntactic}. We consider \tool{} as primarily supporting a bottom-up approach to comprehension with its explanations of expressions; at the same time, its block-level explanations provide some support for top-down comprehension, albeit at the level of dozens of lines of code.

Numerous prior studies have sought to characterize the kinds of questions that arise as programmers read and write code~\citep{Sillito2006, 6227187,ko2007information,sadowski2015developers}. \citet{6227187}, for instance, identified 20 difficult questions that programmers faced when working with unfamiliar APIs. Many of these are answerable using \tool{}. For instance, we have seen \tool{} answer questions like ``what roles do the arguments of a given method play in its usage,'' ``what is the valid range of values for a primitive argument or a given method,'' and ``how do I determine the outcome of a method call?'' from \citet{6227187}'s study, or ``what is the `correct' way to use or access this data structure'' from \citet{Sillito2006}'s study. \tool{} is tuned to answer questions about the high-level behavior of passages of code a couple dozen lines in length, as well as the role of individual expressions in the behavior of a statement.

There exist many methods for evaluating program comprehension, including think-aloud protocols~\citep{von1993program, 10.1287/isre.6.3.286}, memorization~\citep{pennington1987stimulus, 5010283, shneiderman1976exploratory}, comprehension tasks~\citep{5010283}, fMRI readings~\citep{siegmund2014understanding}, near-infrared spectroscopy~\citep{nakagawa2014quantifying}, EEG~\citep{fritz2014using}, and eye tracking~\citep{kather2021through, sharif2010eye, tang2023empirical, al2022readable}. \citet{tang2023empirical} recently observed programmers' gaze when validating and fixing generated code; their study was able to reveal periods of careful attention to generated code. Our own study measures comprehension primarily through a set of questions about generated code, and coarse gaze-based measures of attention.

\vspace{-1ex}
\subsection{Interactive program comprehension aids}
\label{sec:interactive-aids}
The HCI literature is replete with interactive systems research focused on helping people understand programs. Recently-developed research systems have supported the understanding of programs in many senses. For instance, they have supported sensemaking about code~\citep{Catseye2022} and APIs~\citep{Horvath_2022} with novel annotation affordances, understanding of code examples with concept-annotated code snippets~\citep{paralib2022} and collated views of API usages~\citep{glassman2018visualizing,yan2021visualizing}, and understanding how existing web pages are implemented with new methods to inspect their underlying source~\citep{WebCrystal2012, hibschman2016telescope, burg2015explaining}.

Some of this research has, like \tool{}, focused on supporting in-situ understanding of code---that is, augmenting the editor in a way that helps a programmer understand the code that is within their focus. \citet{Hoffswell2018augmenting} introduced a grammar of in-editor visualizations that can be used to understand the values of myriad data types with a small footprint. The projection boxes project~\citep{lerner2020projection} provides a framework for live programming with Python where values of expressions can be viewed adjacent to the lines on which they appear. LEAP~\citep{ferdowsi2023live} applies this idea to support the comprehension of AI-generated code. Tutorons~\citep{head2015tutorons} explores a similar notion to this paper, generating brief ``microexplanations'' for individual lines of code that appear in tutorials to provide on-demand comprehension assistance. A design space of related forms of on-demand help was explored by \citet{potter2022contextualized} in their design of ExplainThis. \tool{} is inspired by this prior work, exploring how some of the above affordances could be brought into the editor to support expression- and block-level code understanding in an era where LLMs are powerful enough to segment and explain code on demand.

A growing number of tools have brought LLMs into the code editor to support code understanding. Notable contemporary examples include CodeHelp~\citep{liffiton2023codehelp}, EasyCode~\citep{EasyCode}, Genie AI~\citep{GenieAI}, GitHub Copilot X~\citep{copilotX}. \citet{nam2023inide} recently designed and evaluated a system of this kind, which supports explanation of selected code, details of API calls used in the code, explanations of domain-specific terms, and provision of usage examples for an API, with a study showing advantages over web browser-based help. \tool{} explores an interaction model where similar kinds of support are provided, in a way that is tightly integrated with the programming assistant to instantly provide clarifying information to programmers without diverting their attention away from the code.
\section{System}
\label{sec:design}
The purpose of \tool{} is to help programmers acquire an understanding of whether generated code matches their intentions, and how they should modify it if it does not. It has a particular focus on helping programmers understand unfamiliar APIs and idioms in generated code. When combined with a programming assistant like Copilot, \tool{} embodies the idea of an \emph{instructive copilot}---that is, a copilot that it explains its work in a way that empowers a user to further refine its output. There is considerable nuance in developing a usable instructive copilot.  A fundamental tension is showing information to programmers in a way that is simultaneously instant and unobtrusive. Below, we describe a set of design goals for instructive copilots that we believe address this tension. The goals are motivated by best practices in interaction design, and by choices that arose during \tool{}'s iterative development. The goals serve to crystallize how our vision departs from contemporary, chat-based approaches to programming help (Section~\ref{sec:interactive-aids}). Specifically, we posit that instructive copilots should provide explanations that are:

\paragraph{D1. Anchored} Programmers should not have to divert their attention from generated code to get assistance in understanding it. \rev{Doing so could induce split attention~\cite{chandler1992split} and undesired cognitive load.} Instead, explanations should appear in-situ, next to the code.

\paragraph{D2. Lightweight}
Explanations should support a basic understanding of the code at a glance. They should be simple enough that they impose only a small load on programmers' working memory. \rev{This is consistent with the principle of minimalism, wherein documentation is kept concise and focused on users' tasks~\cite{carroll1990nurnberg}.}

\vspace{1ex}

\rev{Furthermore, guided by standard usability recommendations in favor of speeding up frequent actions and allowing flexible ordering of tasks~\cite{nielsen1994enhancing}, we recommend that explanations are:}

\paragraph{D3. Easy to invoke}
Programmers should not need to expend any effort to see explanations.

\paragraph{D4. Easy to dismiss}
Explanations should be dismissed automatically when they are no longer needed.

\paragraph{D5. Accessible anytime}
While explanations should be hidden when not needed, it should be easy for programmers to bring explanations back for any generated code when they need them.

\vspace{2ex}

We also posit that explanations should appear at multiple multiple levels of abstraction. As discussed in Section~\ref{sec:comprehension_related}, programmers need to understand the behavior of code not only at the level of individual expressions, but also higher-level structures. Finally, explanations of neighboring expressions should appear in parallel, because the task of understanding a programming statement often requires making sense of the interrelated behaviors of its component expressions. In the upcoming sections, we describe how these goals are addressed in the design and implementation of \tool{}.

\subsection{Interface Design}

\tool{} is designed to deliver on the design goals with the following affordances. We direct readers to Figure~\ref{fig:teaser}, demo video, and the scenario (Section~\ref{sec:scenario}) to see how these affordances appear to users.

\subsubsection{Expression-level explanations}
When the programming assistant suggests a single-line suggestion, \tool{} shows explanations of major expressions that make up that line (Figure~\ref{fig:guidelines}). Major expressions are automatically identified using an LLM (see Section~\ref{sec:implementation-requesting}). Then these expressions are assigned brief descriptive labels. These expression-level explanations are designed to adhere to the design goals, with the following choices:

\paragraph{Anchored (D1)}
To reduce split attention that arises when accessing conventional forms of documentation, explanations appear \emph{anchored alongside the code}, as close to the expressions they explain as possible. Labels are associated with expressions using proximity and shared color on the borders of the expression and label. They appear beneath the suggestion, as we anticipate the context above a line of code will be more important than the context below it when the programmer is choosing whether to accept the suggestion. In the case where many labels are generated, some of the labels are moved further away from their expressions, in which case leader lines are used to associate expressions with labels. The programmer can hover over a label to highlight just that label and underline the corresponding expression. Labels track the code even as the programmer scrolls, zooms, and resizes the editor.

\tool{} does not currently support floating explanations that stand apart from the code. Floating explanations could be useful in situations where there is more to explain that what appears in the code---like other arguments to an API that were not generated. 

\paragraph{Lightweight (D2)} Explanations are \emph{very short}---typically only 1--2 sentences long. Because the explanations are short, programmers have the opportunity to get the gist of an expression very quickly. A consequence of the labels' brevity is that they take up little visual space, allowing programmers to continue to refer to the surrounding code as they consider whether to accept a suggestion.

\paragraph{Easy to invoke (D3)} Explanations are \emph{(near) instantaneous}---they appear within seconds of the appearance of a suggestion (i.e., as quickly as our implementation retrieves them). Appearing automatically, they require no effort to invoke. They also \emph{appear for all expressions in a line at once}. The parallel appearance of explanations can help a programmer visually segment the line into its meaningful parts, and synthesize an understanding of the line by referring to the interrelated meanings of its expressions. 

\paragraph{Easy to dismiss (D4)}
In-situ explanations run the risk of being unwelcome should they distract a programmer. To reduce this risk, we made it easy to dismiss explanations. Explanations are \emph{automatically dismissed} as soon as a suggestion is accepted, or as soon as the programmer clicks away from the suggestion.

\subsubsection{Block-level explanations}
When the programming assistant suggests a long block of code (two or more lines), \tool{} helps a programmer understand the high-level steps taken by the code by showing explanations of blocks of code (i.e., contiguous sets of lines) in the right margin (see example in Section~\ref{sec:scenario}). The interaction design around block-level explanations is the same as expression-level explanations, with the following differences:

\paragraph{Appearance in the margins}
To avoid the risk of occluding code, block-level explanations appear in the right margin. They appear just to the right of the right-most character in the suggested code, or if the suggestion is very long, right after the 80th character. Fading is used on the left side of the explanations to suggest that code is being hidden underneath them. The explanations are associated with lines of code with a single border on the left side of the explanation indicating the extent of code lines to which the explanation applies.

\paragraph{Expression-level drill-down}
When using early prototypes of \tool{}, we found that block-level explanations provided a useful entry point to understand suggestions, but sometimes they did not go into enough detail about what parts of the code did. Therefore, we extended block-level explanations to allow programmers to access details about the code on-demand, allowing them to hover over any line to see expression-level explanations for that line.

\subsubsection{Making explanations accessible anywhere, anytime (D5)}
\label{sec:design-anywhere}

\tool{} was designed to automatically dismiss explanations to reduce their imposition on programmers. A consequence of this choice is that programmers lose access to explanations when they accept code, when they may wish to continue seeing them. \tool{} provides a button that allows programmers to see explanations after they have accepted suggestions. Once they press this button, explanations show not just for the generated code, but for the entire file: block-level explanations are shown in the margins, and the programmer can hover over any line in order to see expression-level explanations. 

\begin{figure}[t!]%
    \centering \includegraphics[width=\linewidth]{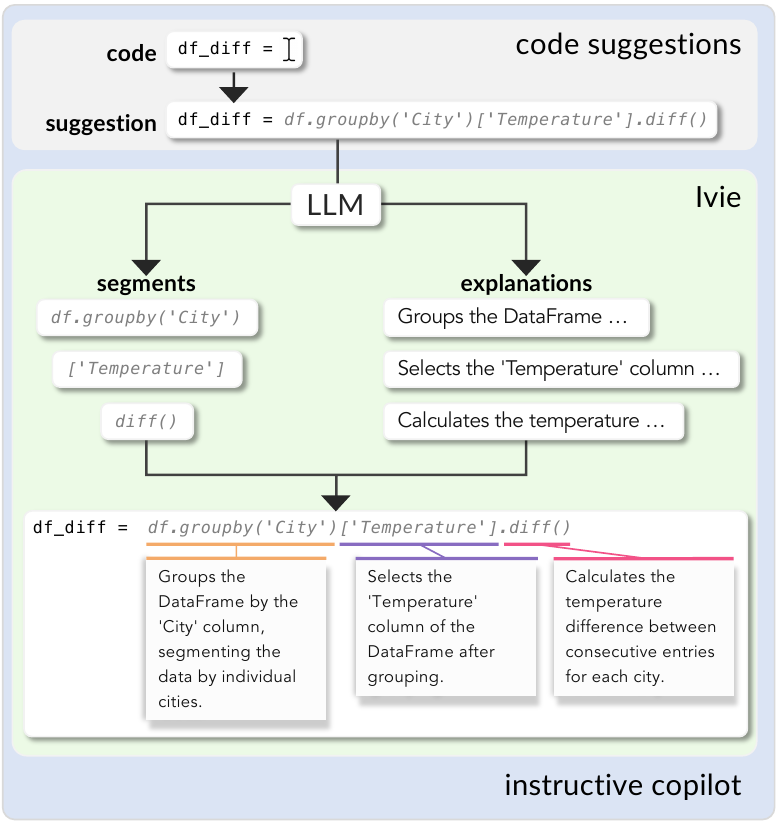}
    \vspace{-2ex}
    \caption{The implementation of an instructive copilot for programming. \textmd{\tool{} creates interactive overlays that explain suggestions made by a programming assistant. When the programming assistant (e.g., Copilot) displays the suggestion, \tool{} submits that suggestion in a prompt to an LLM, requesting that the code be segmented and explained. \tool{} then integrates the explanations into the editor as overlays beneath the expressions they explain.}
    }%
    \vspace{-2ex}
    \label{fig:system}%
     \Description[]{A visual summary of how our implementation takes a code suggestion, uses an LLM to segment and explain it, and then attaches generated explanations to the suggested code.}
\end{figure}

\subsection{Implementation}

\begin{figure}
\includegraphics[width=\linewidth]{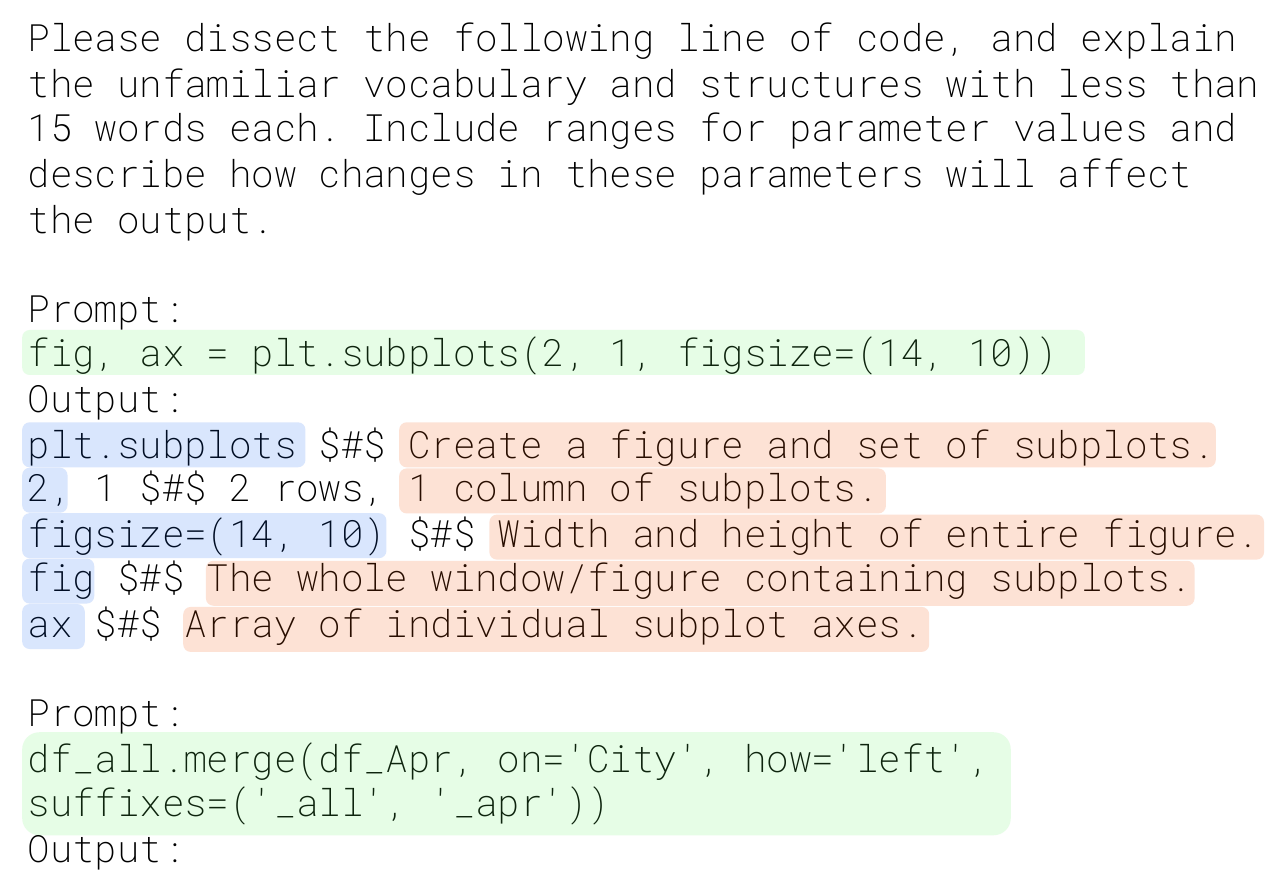}
% \begin{center}
% \begin{tabular}{c} 
% \begin{lstlisting}[style=MySimpleStyle]
% Please dissect the following line of code, and explain the 
% unfamiliar vocabulary and structures with less than 15 words 
% each. Include ranges for parameter values and describe how 
% changes in these parameters will affect the output.

% Prompt:
% fig, ax = plt.subplots(2, 1, figsize=(14, 10))
% Output:
% plt.subplots $#$ Create a figure and set of subplots.
% 2, 1 $#$ 2 rows, 1 column of subplots.
% figsize=(14, 10) $#$ Width and height of entire figure.
% fig $#$ The whole window/figure containing subplots.
% ax $#$ Array of individual subplot axes.

% Prompt:
% df_all.merge(df_Apr, on='City', how='left', suffixes=('_all
% ', '_apr'))
% Output:

% \end{lstlisting}
% \end{tabular}
% \end{center}
% \vspace{-1em}
\caption{A prompt for requesting expression-level explanations of generated code. \textmd{This prompt requests explanations of \textcolor[HTML]{0b6301}{suggested code.} It provides a single example of how it would like code suggestions to be split into \textcolor[HTML]{024cba}{expressions} with accompanying \textcolor[HTML]{be4200}{brief explanations} of those expressions.
}}
\label{fig:prompt}
\Description[]{This figure contains the text of a prompt that Ivie uses to segment generated code and generate explanations of the segments.}
\vspace{-1em}
\end{figure}

A tool like \tool{} can be implemented in a straightforward way, delegating much of the work to external tools and focusing mostly on tailoring their output and tight editor integration. In this section we describe how \tool{} works. A visual summary of \tool{}'s implementation appears in Figure~\ref{fig:system}. From suggestion to explanation, \tool{} undertakes the following steps:

\subsubsection{Activating \tool{}}
\tool{}'s explanations are triggered when a programming assistant---in our implementation, GitHub Copilot---proposes a suggestion. \tool{} listens for suggestions by registering a listener with an internal VS Code API that is triggered whenever ``ghost text'' (VS Code's phrase for an in-editor suggestion) appears in the editor. \tool{} collects the ghost text, and then composes a query to an LLM to retrieve explanations.

\subsubsection{Requesting explanations}
\label{sec:implementation-requesting}
\tool{} then generates explanations. Explanations are composed of two parts:

\begin{itemize}
    \item A segmentation of the code into explainable parts (i.e., blocks or expressions)
    \item Natural language descriptions of those parts
\end{itemize}

Explanations are generated by querying a contemporary LLM. We use the GPT-3.5 \texttt{text-davinci-003} model, and access it through OpenAI's API. The LLM is prompted to simultaneously split and explain the code (see an example in Figure~\ref{fig:prompt}). The prompt asks for explanations to be brief yet informative. The prompt includes a single example demonstrating the intended output.

\begin{figure*}[t!]
\centering
\includegraphics[width=0.7\linewidth]{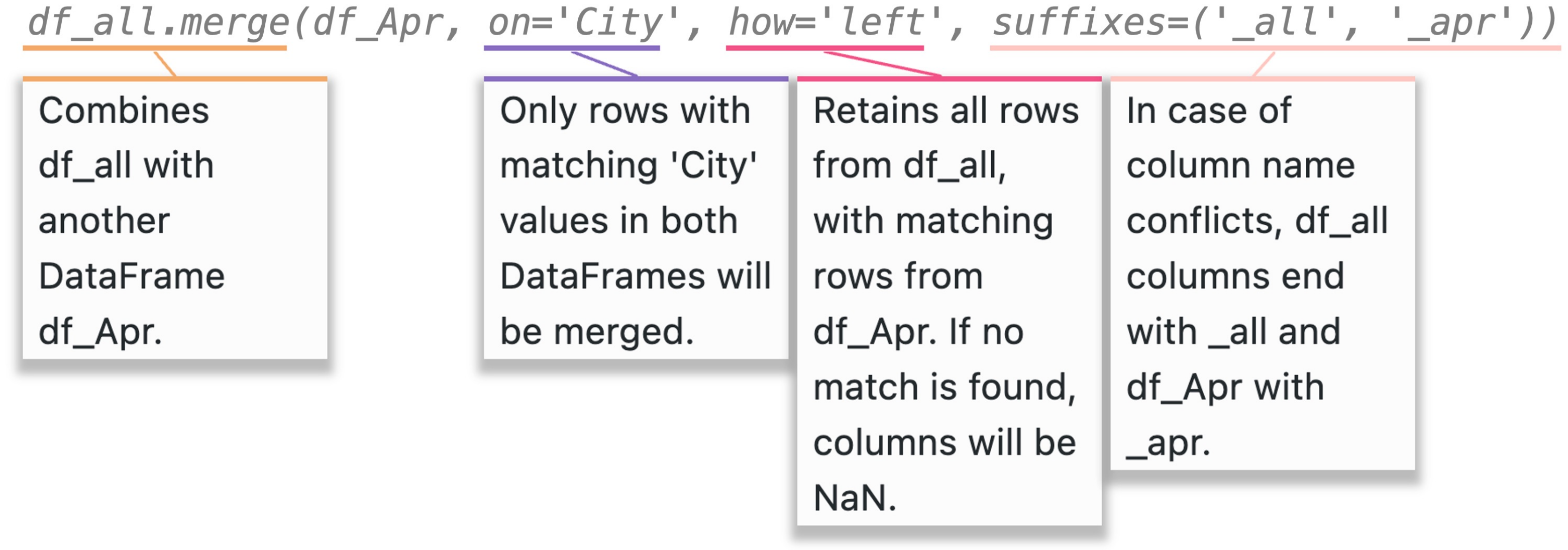}
\vspace{-1ex}
\caption{Explaining expressions. \textmd{After a programming assistant suggests code, \tool{} fetches explanations of that code. If the suggestion consists of a single line, \tool{} reveals explanations of meaningful expressions within that line, such as function calls, and parameters for those calls. The purpose of these explanations is to make explicit the intent of code that may not be self-evident (as might be the case for the programmer seeking to understand the precise behavior of the suggested table merge operation above).}}
\vspace{-1ex}
\Description[]{A piece of code that calls to the “merge” method on the “df_all” object, with arguments that specify what data frame to join on, what column to use for the join, the kind of join action to perform, and what suffixes to assign to merged columns. The “merge” method and the arguments are all explained by Ivie’s explanations. One of the explanations, for the ‘how’ argument, reads “Retains all rows from df_all, with matching rows from df_Apr. If no match is found, columns will be ‘not-a-number.’”}
\label{fig:guidelines}
\end{figure*}

One of two prompts is submitted to the LLM, depending on the size of the suggestion; different prompts are used to generate block-level explanations and expression-level explanations. We set the \texttt{temperature} to $0.5$ and \texttt{max\_token} to $1000$; these parameters were chosen to achieve good explanations with as little latency as possible. When a suggestion consists of two or more lines, \tool{} actually submits many requests: block-level explanations are requested for the full suggested text, and expression-level explanations are requested for all of the constituent lines, in parallel. Prompts and parameters appear in the supplemental material.

Our experiences with the LLM suggests that it achieved performance that was, while not perfect, quite good. In a test we ran of $100$ generated explanations for two API calls used in our usability study (namely, ``\ttt{cv.GaussianBlur(img, (5, 5), 0)}'' and ``\ttt{cv.Canny(img, 100, 200)}''), we assessed 97\% as being correct. Our criteria for correctness were that explanations needed to be complete (i.e., the function name, return value, and all arguments were described), explanations needed to be accurate (i.e., explanations reflected the labeled expressions without any false information), and that the code needed to be properly segmented (i.e., the LLM correctly detected the bounds of each expression without bleeding over into adjacent expressions or delimiters). The positive usability results in Section~\ref{sec:results} suggest that the current error rate around expression segmentation and explanations yields a positive first-use experience. Our discussion elaborates on tensions around using AIs for explaining code in comprehension tools.

\vspace{-1ex}
\subsubsection{Rendering explanations}
Upon receipt of a response from the LLM, explanations are rendered as overlays on top of the editor widget. For expression-level explanations, explanations are rendered all at once. For block-level explanations, we request streamed responses from the LLM (i.e., setting the \texttt{stream} parameter to \texttt{true}). Because block-level explanation requests take longer to fulfill, this allowed us to render explanations for blocks as they are received, rather than waiting for all of them to become available. 

Then, explanations are placed next to the expressions they explain. The default position of an expression label is left-aligned beneath an expression. Label placements are further adjusted to avoid overlap. If two labels overlap, the label to the right is moved rightward until there is no longer overlap. Labels are limited to a maximum width (approximately a few words long) to prevent any one label from taking up too much horizontal space. If a label is rendered far away from its expression (which we encode as less than 50\% horizontal overlap with the expression), leader lines are added to visually link all explanations to their accompanying expressions.

\paragraph{Requesting explanations for the entire file}
When a programmer clicks \tool{}'s button for showing explanations of the entire file's contents (Section~\ref{sec:design-anywhere}), \tool{} submits queries to the LLM as if the entire file's contents were one multi-line suggestion. This has the limitation that when the programmer requests explanations, the explanations might change each time a request is made. This is an artifact of the current implementation; we believe future implementations of \tool{} should preserve explanations between requests.

\begin{figure*}[ht]
\centering
\includegraphics[width=\linewidth]{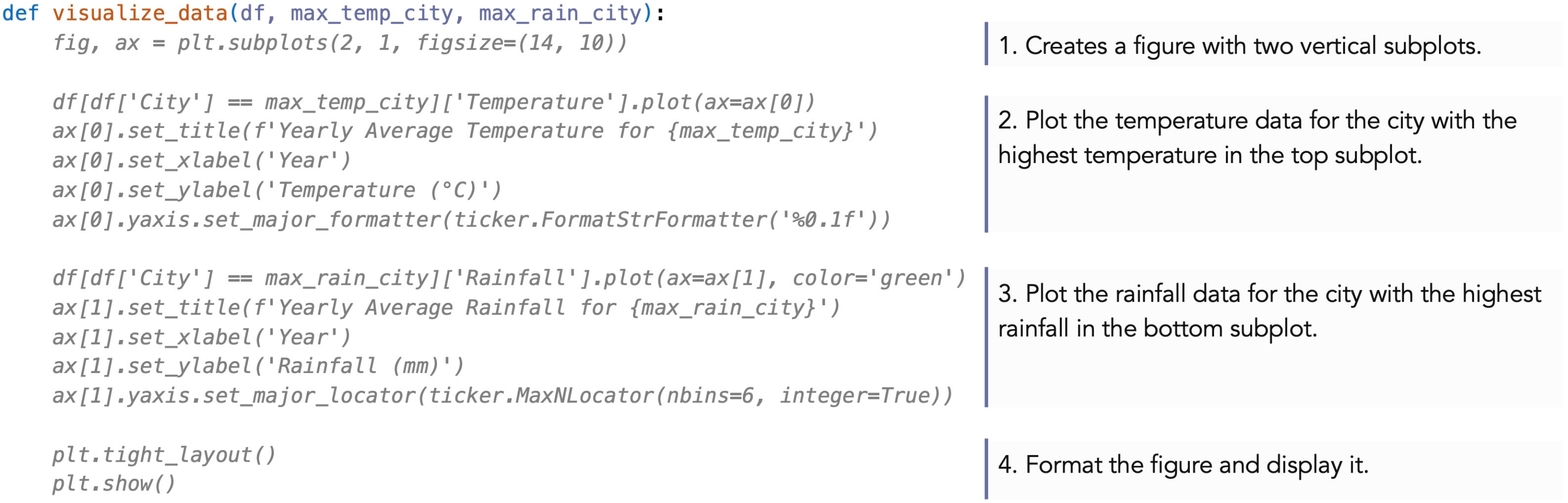}
\caption{Explaining multi-line suggestions. \textmd{When a programming assistant suggests multiple lines of code, \tool{} splits up and explains that code. Its explanations appear in the right margin of the editor. The explanations are meant to help a programmer get a high-level understanding of the code. In the pictured scenario, these explanations might help the programmer understand that the two longest sections of the code suggestion configure each of two subplots, each with a different slice of the data.}}
\vspace{-2ex}
\label{fig:block}
\Description[]{A multi-line suggestion from Copilot, accompanied by Ivie’s block-level explanations. Each explanation describes multiple lines of the suggestion. For instance, the explanation for lines 3-7 of the suggestion is “Plot the temperature data for the city with the highest temperature in the top subplot.”}
\end{figure*}
\section{Scenario}
\label{sec:scenario}

In this section, we describe the experience of interacting with \tool{} in a brief narrative walkthrough. \rev{For a demonstration of interacting with \tool{}, we encourage readers to view the accompanying video figure.} Imagine Dorothy, a climate scientist who is about to perform exploratory data analysis of longitudinal climate data about the Amazon rainforest. Dorothy plans to use a common code-based data analysis toolset: namely, a contemporary code editor equipped with Copilot, and a Python environment with the \texttt{pandas} data manipulation library and \texttt{Matplotlib} visualization library pre-installed. In this scenario, Dorothy is using a code editor that has been extended with \tool{}. Dorothy has passing knowledge of \texttt{pandas} and \texttt{Matplotlib}---enough to do work with them---though she often consults online resources to find out how to fine tune the API functions to manipulate and visualize data as she would like.

\paragraph{Viewing expression explanations}

After Dorothy has spent some time loading, cleaning, and manipulating her data, she decides that she is ready to merge some fields from a derived data frame---\texttt{df\_Apr}---back into the main data frame containing all of her data---\texttt{df\_all}. When she queries Copilot for a suggestion of how to merge the two frames, she sees the following suggestion, augmented with explanations from \tool{} (Figure~\ref{fig:guidelines}).

Without \tool{}, many parts of this line of code would have been difficult to understand. For instance, does a ``left'' join preserve all of the rows from the \texttt{df\_all}, \texttt{df\_Apr}, or both? What values appear in the new columns for rows from \texttt{df\_all} that do not have a corresponding row in \texttt{df\_Apr}? The overlay explanations answer these questions, and others. All rows in \texttt{df\_all} will be maintained; \texttt{NaN} values will be inserted wherever the merged rows have new columns. The explanations remind Dorothy to add a guard to her code to check for \texttt{NaN} values. Dorothy is also reminded that the names of some of the columns will change, as the \texttt{suffixes} argument will rename columns that appear in both data frames. In this way, Dorothy acquires a detailed idea of the result of the merge that otherwise would need to have been pieced together from selective reading within the Pandas documentation. 

\paragraph{Viewing high-level explanations of code suggestions}
Later, Dorothy arrives at a stage of analysis where she would like to visualize a slice of her data. She would like to examine the temperature and rainfall for the cities that have experienced each most extremely. She outlines a Python function for visualization, and then Copilot makes a 16-line suggestion. \tool{} augments the suggestion with descriptive labels (Figure~\ref{fig:block}).

The explanations help Dorothy orient to the code. Her initial skim of the code leads her to erroneously interpret that the function sets up one figure, and then configures the x-axis (in the sequence of lines beginning with ``\texttt{ax[0]})'', and then the y-axis (in lines beginning with ``\texttt{ax[1]}''). She pauses to examine the labels that \tool{} provides in the right margin, and by the time she reads the third label, she realizes that the blocks of code she associated with the x- and y-axes in fact configure two distinct plots. Now she knows that she can edit the block of code for ``\texttt{ax[0]}'' to configure the plot for the temperature data for the city with maximum temperature. She further validates her understanding of what individual lines of code do by hovering over them---as she hovers over each line, expression-level explanations show for that line. These explanations allow Dorothy to understand how she can configure the formatting string ``\texttt{\%0.1f}'' to include more significant digits, and that she can use the ``\texttt{integer}'' parameter to configure whether ticks in the y-axis of the second plot are constrained to integer values.
\section{Study Design}
To evaluate \tool{}, we conducted a usability study. The study focused on the impact of \tool{} on working with generated code using unfamiliar APIs. We sought answers to the following questions:

\paragraph{Q1. Does \tool{} improve understanding of generated code?} 
\tool{}'s goal is to explain generated code to support high-level understanding, so we evaluated programmers' understanding of generated code.

\paragraph{Q2. Does \tool{} influence how much attention programmers give to generated code?} 
Our hypothesis was that programmers would more closely examine generated code when it was accompanied by lightweight explanations.

\paragraph{Q3. How distracting is \tool{}?} 
While \tool{} was designed to only minimally distract programmers, we sought evidence of just how much distraction they really experienced.

\paragraph{Q4. How does \tool{} compare to chat-based AI code comprehension aids?}
What are the benefits and downsides of \tool{} compared to other contemporary alternatives for comprehension assistance?

\subsection{Participants}

32 programmers were recruited from academic mailing lists in the computer science department at the University of Pennsylvania. 31\% were doctoral students, 63\% were master's students, and 3\% were bachelor's students.\footnote{Some percentages describing participant backgrounds do not add up to 100\%. This reflects occasional non-response to questionnaire items.} 16\% of participants reported their level of skill with Python to be advanced, and 50\% reported proficient, 31\% beginner, and 3\% no experience.\footnote{All proficiency questions allowed participants to report the level of ``expert.'' No participants selected this level for any question.}

As intended, programmers were largely unfamiliar with the libraries involved in the programming tasks.  The main library used in the tasks was the OpenCV computer vision library~\cite{OpenCVOp4:online}. 38\% of programmers reported no experience with this library, 53\% were beginners, and 9\% were proficient. When asked how familiar they were with computer vision generally, 31\% were not at all familiar, 53\% a little familiar, 13\% somewhat familiar, and 3\% very familiar. Participants also had very little experience with the library used in the open-ended programming task (see Appendix~\ref{apx:game}).

31\% of programmers had previously used Copilot. 90\% of the programmers who had used Copilot reported using it for a few months or less, and only one participant reported using it for nearly a year (as reported in a multiple-choice question). Although these participants had previously used Copilot, we anticipated they would not experience a bias in the form of a novelty effect with \tool{}, because our baseline was also novel in extending Copilot's functionality. Because our baseline involved interacting with an LLM-based chatbot, programmers were also asked about their experience using LLM-based chatbots. All (100\%) participants had previously used ChatGPT before, and 59\% had additionally used some other chat-based AI. 38\% reported having a few months of experience using chat-based AI, 47\% had about half a year, 13\% had about one year, and 3\% reported between one and two years of experience.

\subsection{Baseline}

Our study compared \tool{} to a strong modern baseline. The baseline was the modern VS Code IDE with its built in documentation tooltips, access to a web browser, and an editor plugin providing an AI-based chatbot. We call our baseline condition the ``chat condition'' due the presence of the chatbot.

We chose a chatbot plugin for the baseline to represent a family of recently-developed AI-based code comprehension plugins (e.g.,~\cite{copilotX,GenieAI,CodeiumA7:online}). In these plugins, the main feature is a chat window to left side of the editor where a programmer can ask questions about their code, and follow-up questions after the chatbot responds. These plugins often support the ability to select arbitrary code and request an explanation of that code from the editor's menus. We chose the plugin ``EasyCode''~\cite{EasyCode} which provides all these features and has lower latency among alternatives.

To promote parity between the baseline and \tool{}, we configured both the chatbot and \tool{} to use GPT-3.5 for their explanations. Aside from the presence of the chat functionality, the editor was configured identically for both the baseline and \tool{}.

\subsection{Procedure}
Each participant came to our lab for a one-hour-long session. To reduce demand characteristics~\cite{orne2017social} that might have biased a participant in favor of \tool{}, we told them our goal was to understand the influence of two explanation tools---both the chat baseline and \tool{}---on understanding generated code. To avoid leaking our role in developing \tool{}, we referred to both tools using pseudonyms ``chat explanations'' and ``overlay explanations.'' Participants consented and completed a questionnaire about their programming background before completing the following stages:\footnote{All questionnaires, task instructions, starter code, and assessments can be viewed in the supplemental material.}

% Each participant consented and completed a questionnaire about their programming background before completing several stages. The materials for all of these stages can be found in the supplementary materials. [You might also want to mention what specific materials can be found in the supplemental materials.]

\subsubsection{Tutorial}

The programmer was instructed in the use of all tools used in the study---Copilot, the chat baseline, and \tool{}. The tutorial consisted of a 5-minute slide presentation, and an activity the programmer was guided to create a data visualization with Copilot and invoke both the baseline and \tool{} to access explanations. 

\subsubsection{Timed Programming Tasks}

The programmer then undertook two timed programming tasks (Tasks~A and B), one with \tool{}, and one with the baseline, with order of tasks and interfaces counterbalanced. Each task required the programmer to write a short snippet of image manipulation code using OpenCV. As a prompt, participants were provided an input image, a target output image, a goal (e.g., ``blur the image so it resembles the target image''), and starter code. Both tasks were designed to be similar in complexity and focus. Each task require understanding of an OpenCV API that would be unfamiliar to the programmer. The tasks were validated through extensive piloting until we were confident that programmers would almost always be recommended the expected APIs by Copilot, and that the explanations from \tool{} would be coherent. Tasks lasted 5 minutes each. This amount of time was sufficient to ensure that all programmers would make some progress (e.g., achieve some blurring of an image), while introducing a cutoff that let us compare how closely each programmer approached to the target parameters across conditions. Programmers were allowed to use a web browser in either condition, though no programmers did so.

After each task, the programmer reported task difficulty using the NASA-TLX questionnaire~\cite{NE08F1443:online} and answered Likert scale questions about how useful the available tools were in helping them understand generated code. To decrease the likelihood that programmers studied code to an unnatural degree during the programming tasks, we had them complete both tasks before telling them that their understanding of code would be assessed.

\subsubsection{Timed Comprehension Assessments}
\label{sec:comprehension}

After each programming task, comprehension of generated code was assessed with a timed assessment. The assessment focused solely on the OpenCV API call---i.e., what we believed would be unfamiliar generated code---that was used for image manipulation in the programming task. Each assessment consisted of 20 ``yes'' / ``no'' questions about the API call, including the function name and its arguments (see details in Appendix~\ref{apx:timed}). Our choice to time the assessment was inspired by priming tests that have been used in program comprehension (e.g.,~\cite{pennington1987stimulus}), where response time in answering questions is used to measure the strength of learned associations. A potential threat to validity is that answers to comprehension questions resembled \tool{}'s in length (i.e., programmers indicated the meaning of an expression by selecting from a set of short text descriptions). We note, however, that answers resembled descriptions of expressions similarly well for the the baseline and \tool{}. After running the study, a follow-up investigation of 12 randomly-sampled programmers' session videos showed that, of the correct answers, $83.3\%$ exactly matched the description of the expression from \tool{}, and $88.9\%$ exactly matched the description of the expression shown by the baseline. The difference is that descriptions of expressions appeared as part of much longer texts in the chat baseline, which is just the problem that \tool{} is meant to address.

\subsubsection{Open-Ended Programming Task}

The programmer was given the remainder of the time (typically 10 minutes) to explore \tool{}'s support for an exploratory programming task (see details in Appendix~\ref{apx:game}). This task was designed to help participants ground qualitative feedback on the anticipated usefulness of the tool.

\subsubsection{Questionnaire + Interview}
The programmer filled out a questionnaire where they reflected on the usability of \tool{} and the baseline. If there was time remaining in the session, we conducted a brief semi-structured interview.

\subsection{Measures}

To answer our research questions, we measured the following:

\emph{1. Code comprehension}. We measured programmers' accuracy and speed in replying to comprehension questions.

\emph{2. Attention}. We measured attention as the amount of time a programmer fixated on generated code. A Tobii Pro Spark eye tracker~\cite{Enterthe17:online} was used to collect gaze position during all of the programming tasks. The code editor was instrumented to log the positions of all generated code, as well as the baseline's chat window, and all of the labels that \tool{} showed. The eye tracker was calibrated before each of the programming tasks.\footnote{With the exception of the first five participants, for whom the eye tracker was calibrated only once at the beginning of the first timed programming task.} 

\emph{3. Distraction}. Programmers  answered Likert scale questions about how distracting they found \tool{}'s expression explanations, its block explanations, and the chat baseline. They reported task load using the NASA-TLX index.

\emph{4. Comparisons to baseline}. Programmers answered Likert scale questions comparing \tool{} to the baseline, and were asked to elaborate on the comparative benefits of the two tools.

\subsection{Analysis}

All comprehension questions were assessed using linear mixed-effects models. These models incorporated the tool, the order of tools, task order, and interactions as fixed effects, and participant ID as a random effect. \rev{Statistical significance was assessed using an F-test with Satterthwaite’s estimate of effective degrees of freedom~\cite{satterthwaite1946approximate}, with the Holm-Bonferroni method~\cite{holm1979simple} to correct $p$-values.} For comparisons of Likert scale responses, we assess significance with a two-tailed Wilcoxon signed-rank test~\cite{wilcoxon1992individual}. For all tests, the threshold for statistical significance was $\alpha=0.05$. Qualitative themes from questionnaires and interviews were determined following a thematic analysis process~\cite[Chapter 5]{blandford2016qualitative}, wherein one author performed an initial open coding and axial coding pass, a second author revised the complete results, and then the first author validated and made slight adjustments the revised results before writing reports that appear below.

\section{Results}
\label{sec:results}
\subsection{RQ1. \tool{} improves code understanding}
\label{sub:understand}

\paragraph{Comprehension questions}

\begin{figure}[t!]%
    \centering
    \includegraphics[width=\linewidth]{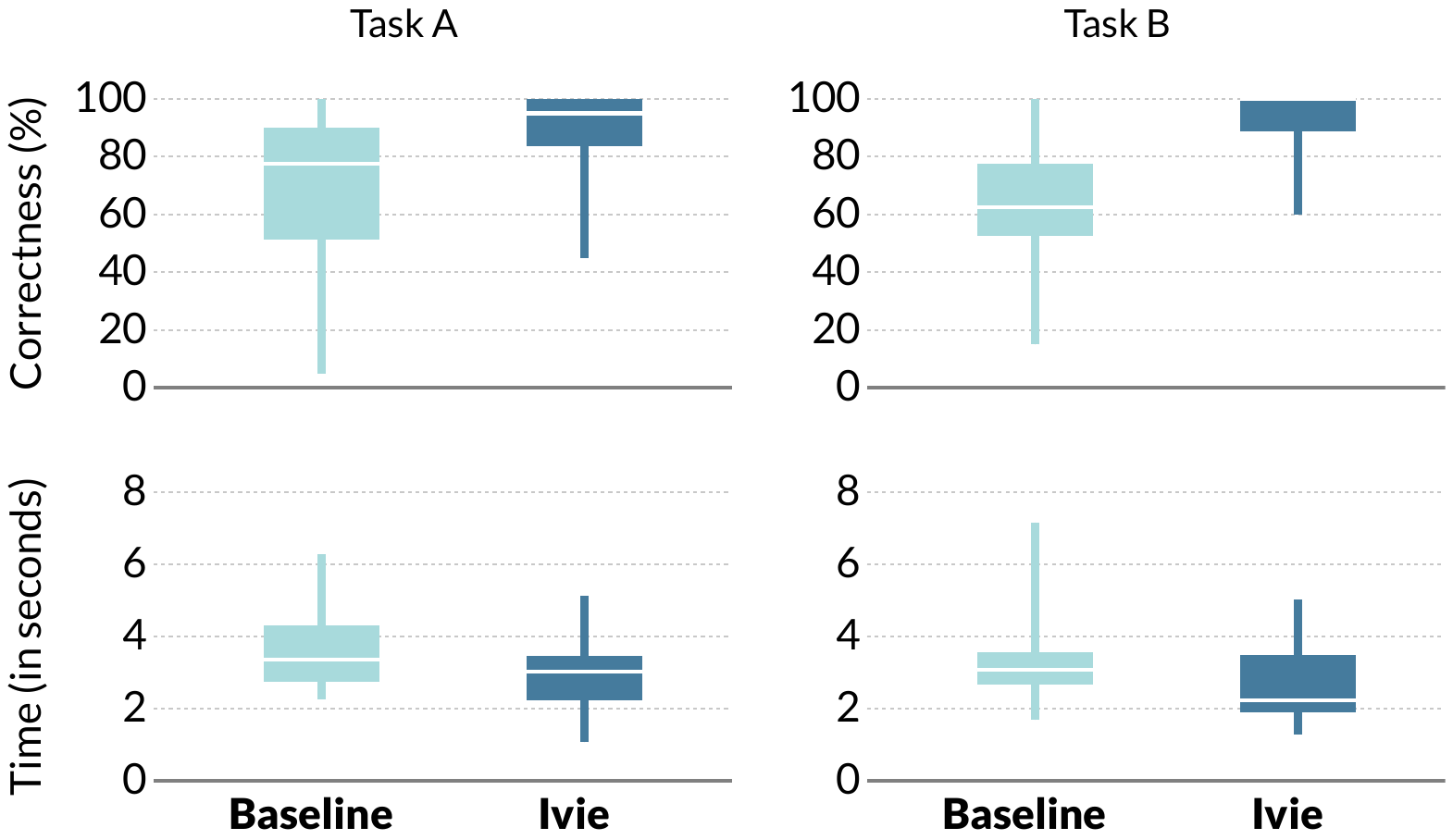}
    % \vspace{0.1in}
    \vspace{-4ex}
    \caption{Comprehension results. \textmd{Each data point for a box plot corresponds to a programmer's score on the assessment or their average question response time. Data is clustered by which tool (\tool{} or baseline) was used for the programming task (Task A or B) associated with the assessment. For correctness, higher values are preferred. For time, lower values are preferred. The differences of \tool{} vs. chat baseline on both correctness ($p<0.001$) and time ($p = 0.011$) are statistically significant.}}
    \label{fig:box}%
     \Description[]{Four box-and-whiskers compare the comprehension of participants using Ivie and chat baseline in tasks A and B, measured using correctness and time to answer. Participants consistently answered more questions correctly, in less time, when using Ivie instead of the baseline.}
    \vspace{-2ex}
\end{figure}

Programmers answered comprehension questions significantly more correctly for tasks completed with \tool{} than with the baseline (see Figure~\ref{fig:box}) ($F = 23.6$, $p < 0.001$). When using, \tool{}, they answered an average of $90.2\%$ of questions correctly, in contrast to $65.0\%$ with the baseline ($\sigma = 26.7\%$). Programmers also answered questions more quickly with \tool{} ($F = 9.82$, $p = 0.011$), answering questions in an average of $2.8$ seconds ($\sigma = 1.0$) about code they had seen in the \tool{} condition, versus $3.6$ seconds ($\sigma = 1.3$) for the baseline condition.

\paragraph{Self-reported understanding}

In their Likert scale feedback, programmers agreed that they understood the code when using \tool{} ($median = 7$ out of 7 on a Likert scale, $\sigma = 0.92$), and found the explanations helpful for clarifying the code ($median = 7$, $\sigma = 0.87$). They reported significantly higher agreement than for the baseline for both questions ($median = 6$, $\sigma = 1.69$, $W = 22$, $p = 0.002$, and $median = 5.5$, $\sigma = 1.94$; $W = 26$, $p = 0.001$) (see Figure~\ref{fig:stacked_2}).

\paragraph{Task progress}
An indirect measure of code understanding is programmers' ability to successfully perform the programming task. We observed that 4 programmers ended a task in an error state with the baseline, whereas no programmers did so in the \tool{} condition. Appendix~\ref{apdex:a} further discusses that, for 2 of 4 configurable parameters in the APIs under study, programmers arrived significantly closer to the target values.

\subsection{RQ2. No observed effect of \tool{} on attention}

\paragraph{Duration of attention on generated code}
Programmers spent less time less time looking at generated code in the \tool{} condition ($\mu = 3.36\,\textrm{minutes}$, $\sigma = 1.56$) in comparison to chat baseline (chat baseline: $\mu = 4.13\,\textrm{minutes}$, $\sigma = 1.59$). However, \rev{the test did not identify this difference as statistically significant.} ($F = 4.25$, $p = 0.14$).

\paragraph{Self-reported effect on attention}

Some programmers reported that \tool{} influenced the way they looked at code when answering the open-ended questions, and in particular that they more closely examining the generated code with \tool{} (P12, P31, P32). P32, for instance, told us that ``\textit{when using [\tool{}] , I carefully examine the completions instead of quickly accepting them.}'' P12 described themselves as ``checking everything'' when using \tool{}. Programmers sometimes felt that \tool{} encouraged the behavior of carefully examining code (P1, P26).

\subsection{RQ3. \tool{} is not (too) distracting}

\begin{figure}[t!]%
    \centering
    \includegraphics[width=\linewidth]{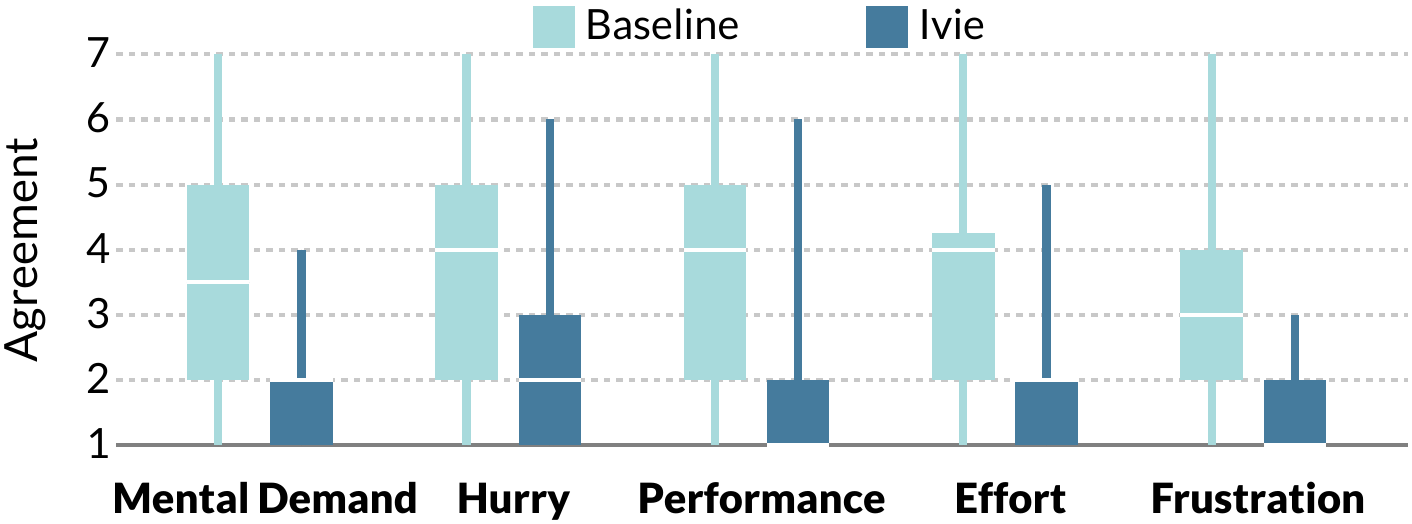}
    \vspace{-3ex}
    \caption{Task load results. \textmd{Shown are programmers' responses to 5 items from the NASA Task Load Index, collected after each timed programming task. Responses are grouped by which tool was used in the task. For all items, lower values are preferred.}
    }%
    \vspace{-5ex}
    \label{fig:nasa}%
     \Description[]{Box-and-whiskers plots comparing responses to the task load questionnaire with Ivie and the baseline. In every comparison, the 1st, 2nd, 3rd, and 4th quartiles show lower task load for Ivie than for the baseline. Dimensions of task load that were measured included mental demand, hurry, performance, effort, and frustration.}
    \vspace{2ex}
\end{figure}

\paragraph{Task load}
Programmers reported task load following each programming task. Task load was assessed using five dimensions from the NASA task load index: mental demand, hurry, performance, effort, and frustration. On all dimensions of task load, \tool{} was seen as imposing less load than the baseline, including mental demand ({\tool}: $median = 2$, $\sigma = 0.92$ vs.  baseline: $median = 3.5$, $\sigma = 1.62$; $W = 4.5$, $p<0.001$), hurry  ({\tool}: $median = 2$, $\sigma = 1.23$ vs. baseline: $median = 4$, $\sigma = 1.61$; $W = 10$, $p<0.001$), performance ({\tool}: $median = 1$, $\sigma = 1.15$ vs. baseline: $median = 4$, $\sigma = 1.70$; $W = 40$, $p<0.001$), effort ({\tool}: $median = 2$, $\sigma = 1.02$ vs. baseline: $median = 4$, $\sigma = 1.56$; $W = 16$, $p<0.001$) and frustration ({\tool}: $median = 1$, $\sigma = 0.56$ vs. baseline: $median = 3$, $\sigma = 1.79$; $W = 0$, $p<0.001$) (see also Figure~\ref{fig:nasa}).

\paragraph{Self-reported distraction}

\begin{figure*}[t!]%
    \centering
    \includegraphics[width=\linewidth]{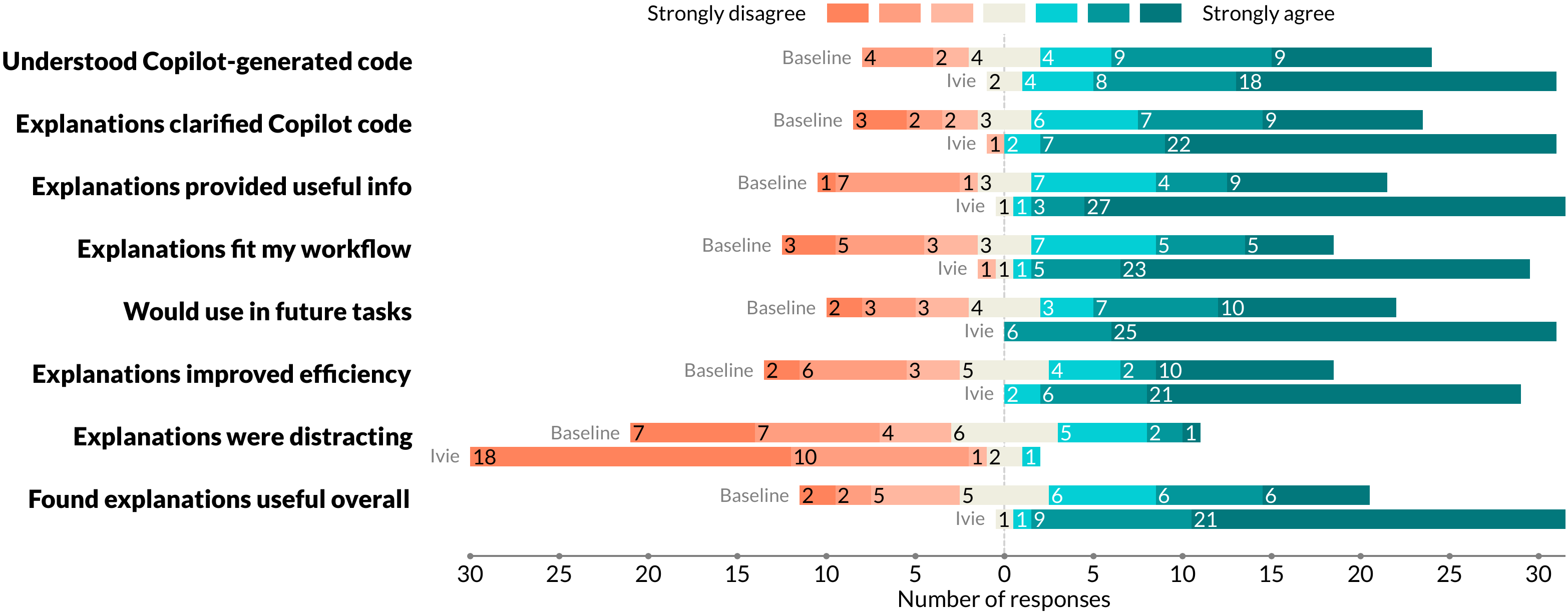}
    \vspace{-4ex}
    \caption{Self-report reflections on tool usability following timed programming tasks. \textmd{Shown are programmers' responses to 8 self-report questions asked on a 7-point Likert scale after each timed programming task.
    }}%
    \label{fig:stacked_2}%
     \Description[]{Stacked bar charts of Likert scale responses to questions about the utility of Ivie and the baseline. Distributions skew in favor of Ivie for how much it helped participants understand code, the usefulness of explanations, participants’ willingness to use it in future tasks, and perceived efficiency. Distributions skew towards suggesting the baseline was seen as more distracting.}
    \vspace{-2ex}
\end{figure*}

After each task, programmers were asked to report how distracting they found the explanations provided by the tool in that task. \tool{}'s explanations were reported as not distracting ($median = 1$ out of 7 on a 7-point scale, $\sigma = 1.01$), and significantly less distracting than the explanations from the chat baseline ($median = 3$, $\sigma = 1.72$; $W = 30$, $p < 0.001$, also see Figure~\ref{fig:stacked_2}). In the closing questionnaire, programmers reported that \tool{} was less overwhelming ($median = 2$, $\sigma = 0.94$) and distracting ($median = 2$, $\sigma = 1.02$) than the chat baseline (where 1 indicated that the chat baseline was worse, and 5 indicated that \tool{} was worse, also see Figure~\ref{fig:stack}). 

\subsection{RQ4. \tool{} complements chat-based AI code comprehension aids}
\label{sec:results-desirability}

\begin{figure*}[t!]%
    \centering
    \includegraphics[width=.97\linewidth]{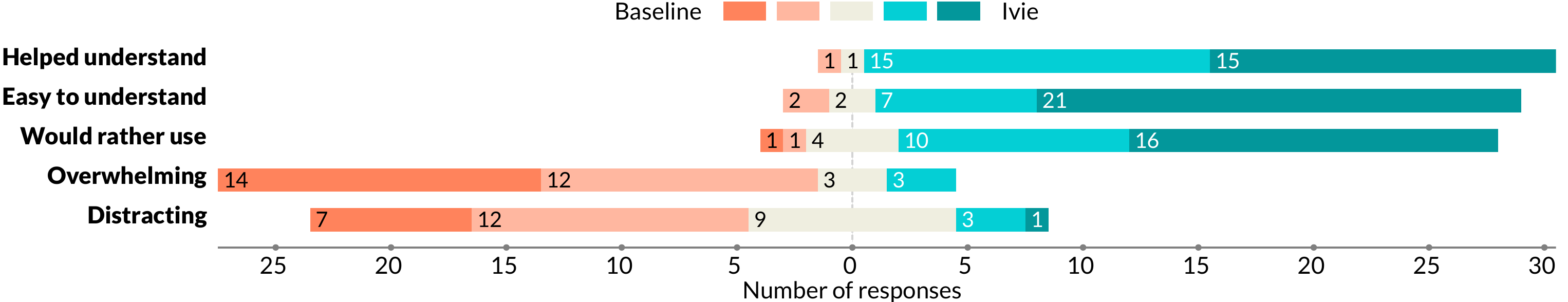}
    \vspace{-1ex}
    \caption{Summative self-report reflections. \textmd{Programmers compared \tool{} to the baseline on a 5-point ordinal scale according to five different dimensions of usability (see labels on left). Pictured is an overall consensus that \tool{} better helped programmers understand generated code, \tool{}'s explanations were easier to understand, programmers would rather use \tool{} for future tasks involving AI-generated code, and the explanations from the chat baseline were more overwhelming and distracting.}}%
    \label{fig:stack}%
     \Description[]{Stacked bar charts of Likert scale responses to questions directly comparing Ivie and the baseline. Distributions show a preference for Ivie in its ability to help participants understand code, and overall preference; the baseline was reported as more overwhelming and distracting.}
    \vspace{-2ex}
\end{figure*}

\paragraph{Direct comparisons to baseline}
As mentioned above, programmers reported \tool{} as being significantly less distracting and overwhelming than the chat baseline, and reported better understanding the code. When comparing the two tools on a 5-point Likert scale (with 5 indicating a preference for \tool{}) programmers also reported that \tool{}'s explanations were clearer ($median = 5$, $\sigma = 0.87$). Qualitative feedback painted a picture of \tool{} as complementary to tools like the chat baseline. In the words of P3, they were both ``great tools to use, both have their utilities. Hard to choose only one out of them. Would be helpful to use in conjunction.''  Discussion Section~\ref{sec:validating} talks at length about the perceived benefits of \tool{}, including the conciseness and anchored nature of the explanations. Perhaps for these reasons, participants largely indicated that if they were to choose one of the two tools for future tasks involving AI-generated code, they would prefer \tool{} ($median = 4.5$ of 5, $\sigma = 0.99$) (see also Figure~\ref{fig:stack}). The comparative advantages of the chat baseline were its support for acquiring conceptual understanding for a task (P22), and deciding what code to write (P6, P26), level of detail in explanations (P3, P29, P32), and the ability to ask follow-up clarification questions (P5, P8, P10, P21). The ability to ask follow-up questions seemed a particularly resonant feature---22 of 32 programmers indicated in their questionnaires that the ability to ask follow-up questions of the chat baseline was either somewhat or very useful.

\section{Discussion and Future Work}
\label{sec:discussion}

\subsection{Limitations}
Our study findings are limited in the following ways. First, our results represent usability among a limited sample of the broader programmer population. The participants in our study were primarily master's and doctoral students. About two-thirds of participants had not used Copilot before the study. We anticipate that experiences with \tool{} would vary for programmers who are less experienced at programming (and so less able to progress generally), more experienced (and so more knowledgeable about their toolsets), and those who have more established workflows of writing code with programming assistants (and therefore perhaps more resistant to extensions to those programming assistants).

Second, our metric of fixation time does not necessarily capture the aspects of attention that are important. We measured attention as the total amount of time spent looking at generated code. This is a coarse-grained measure, in that it is does not differentiate between desirable attention---like the first read-through of code---from undesirable attention---like time spent debugging code that was not properly understood. In retrospect, we note that more nuanced measures may be necessary to assess whether generated code is attracting the kind of attention it should.

Finally, the study only examined a very limited subset of tasks. The tasks were narrow, focusing on the understanding of individual APIs in short generated programs. To assess the utility of \tool{} in supporting programming practice more broadly requires evaluation on a broader set of tasks of various domains and levels of complexity.

\subsection{Design implications for instructive copilots}

From our design and study, what do we now know about the effective design of instructive copilots?

\subsubsection{Reexamining the design goals}
\label{sec:validating}

A first question is: are the design goals we posed in Section ~\ref{sec:design} useful guides for instructive copilots? Qualitative feedback provides validation for these goals as useful guides for instructive copilots in this domain:

\paragraph{D1. Anchored.} One of the benefits of \tool{} was that its explanations were ``visually accessible'' (P15). Participants appreciated that they were targeted to specific places in the code (P6) and ``dissected the parameters.'' (P4) This stood in contrast to chat-based help, which was seen as giving ``long paragraphs of general ideas'' (P26) and requiring one to move their ``sight outside of my code editor, which was pretty annoying.'' (P11).

\paragraph{D2. Lightweight.} Another frequently-mentioned advantage of \tool{} was the lightweight nature of the explanations. Participants appreciated the conciseness of explanations (P4, P6, P15, P18, P22), describing them as easier to understand (P22, P25), simpler (P23), and less overwhelming (P4). In contrast, the chat explanations were often seen as providing too much information (P2, P6, P9, P24).

\paragraph{D3. Easy to invoke.}
Participants appreciated that \tool{}'s explanations appeared instantaneously (P11, P19), and in particular that  the explanations appeared right after Copilot generated the code (P12, P28). One questionnaire item asked participants how useful they found the ability to receive explanations instantaneously; 20 of 32 participants reported it to be at least somewhat useful.

\paragraph{D4. Easy to dismiss.} Participants generally did not comment on the dismissal mechanism. To us, this indicates that for most participants the design was unremarkable and fit adequately. That said, some participants wished explanations were always-on (i.e., never dismissed) (P17, P24, P26), and P32 wished for a toggle button to turn explanations on and off.

\paragraph{D5. Accessible anytime.} Following the \tool{} task, 25 of 32 participants conveyed that it was at least somewhat useful to access explanations at any time. Some participants (P19, P26) conveyed that they would have liked if there was even less friction to bring explanations back up.

\subsubsection{Expanding the design goals}

Our study also revealed opportunities to extend our notion of the instructive copilot. We pose three additional design goals, following participant feedback. Namely, generated explanations should be:

\paragraph{D6. Expandable.} Explanations should let programmers ask for more details.
Several participants wished for the ability to expand explanations (P12, P18, P20). As envisioned by P12, ``ideally, everything starts with a short explanation. If I don't understand something, I could click for more details on that parameter.'' One way to expand explanations is to let programmers submit follow-up questions about explanations to their AIs (P6, P8, P11, P21, P23).

\paragraph{D7. Adaptable.} Explanations should be adapted to the programmer.
Some participants desired that \tool{} adapted explanations to them (P11, P28). Explanations were seen as unnecessary for code that programmers already familiar with (P27, P30, P31). When an explanation did not convey any useful information, it was suggested that the explanation was not shown (P6, P31); for instance, P6 singled out one such explanation, saying that ``Descriptions like `object' are unhelpful. Filtering such terms would improve [Ivie's explanations].'' Future instructive copilots could be selective about what is explained and how, if they were extended to have a reasonable notion of programmer knowledge and needs.

\paragraph{D8. Controllable.} It should be possible for users to configure what content is explained and how it is explained. Some participants asked for controls that allowed them to influence the level of detail in explanations (P5, P22). They also desired control over the granularity at which code was explained (i.e., at the expression- or block-level) (P19, P22). Furthermore, some wished for the ability to request explanations for specific code selections (P6, P17, P18).

\subsubsection{Bringing instructive copilots outside of the code editor}

While in this paper we validate the idea of the instructive copilot as an augmentation to the code editor, we envision that instructive copilots could be useful in other AI generation settings as well. In particular, an instructive copilot could be useful in any setting where there is a timely opportunity for a user to learn more about AI-generated content. For example, perhaps a digital artist would wish to better understand how they reproduce a visual effect that was performed by an AI. For that artist, an instructive copilot might annotate the graphical objects it edited, and reveal the sequence of tools that could be used to attain that effect manually. Another example is writer who is working with AI to compose a passage that is full of references to external work. Perhaps the instructive copilot would describe the nature of the referenced work for any of the references that it generates of which the writer is unfamiliar. We suspect that all such settings would benefit from the same design goals of anchored, lightweight explanations that are easy to invoke and accessible anytime. Where we believe there is interesting variation to explore between applications is in three aspects of an explanation:

\paragraph{Why are explanations needed?} For programming assistants, instructive copilots are useful for helping people understand unfamiliar parts of generated code. In other applications, we foresee several reasons to explain AI-generated content. The first reason is that the generated content is confusing, as in the case of unfamiliar code or, say, generated math notation. A second reason is if a user wants to verify AI-generated content, as is the case of a writer who wishes to check a set of generated citations, or follow along with an AI-generated proof. A third reason is to know how to reproduce AI-generated content, as is the case of the artist who wants to know how an AI achieved a visual effect. The reason for explanation influences the next decisions of what gets explained and how. 

\paragraph{The unit of explainable material} In the context of programming assistants, we foresaw that programmers would need to understand individual expression (i.e., components of API calls) and blocks of code (e.g., multi-line idioms). In other scenarios, users would need explanations of generated graphical objects and citations; of texts at the level of individual words (like jargon), phrases (claims), full passages (chains of reasoning), and even multimedia content like segments of audio and video (e.g., effects produced in creativity tools). There needs to be a mechanism for segmenting AI-generated content (as we do with an external LLM) to identify explainable units and allows them to serve as anchors.

\paragraph{The explanation} The content of an explanation follows from the purpose of explanation. If an explanation is meant to clear up confusing material, it might provide generated descriptions or links to external references. If it is instead meant to aid in the verification of AI-generated content, it might instead link to additional documents that provide supporting or conflicting information. And if it is meant to help someone reproduce AI-generated content, then it might provide procedures for doing so. In the case of programming assistants, we recommended that explanations be extremely concise, in part because the explanations would be interspersed between code generations that may be just seconds apart. In other settings, it may be less necessary for explanations to be concise, particularly if they are providing complex external information that may be supportive in verifying AI-generated content.

\paragraph{Dismissal mechanism} Our design makes use of automatic dismissal when a user clicks away from generated content. This may not be the only appropriate choice. In other cases, where a user wants to continually refer back to the explanation (as in a writer who may want to continue looking at clips from a cited document), it may be better to preserve explanations until the user moves their focus away from the current paragraph or section. In other circumstances, it may be appropriate for explanations to be always-on, as was the desire of some programmers in our study.

\subsubsection{A critique of using AI to explain AI-generated content}
The possibility of prototyping a tool like \tool{} has only recently become possible with the release of contemporary AI tools. As with all AIs, those used by \tool{} make mistakes. If the AI for \tool{} makes a mistake, a programmer may draw inappropriate conclusions about the behavior of their code. In the worst case, this could lead to significant bugs and negative side effects in the generated code. In less severe cases, it could lead the programmer to reject useful suggestions, or slow them down as they attempt to comprehend code. In this way, one of the motivations of this tool---to better inform users about their AI-generated content---is undermined in part because AI-generated explanations may be themselves incorrect.

We see these potential downside as further motivating research in AI to produce validated texts. In the meantime, \tool{} might still be deployable in settings where they provide value despite inaccuracies. Programmers already use tools like ChatGPT~\cite{ChatGPT40:online} as a code understanding aid. Conventional documentation itself contains inaccuracies and outdated information~\cite{robillard2011field,uddin2015api}, and programmers adjust to this reality. Amidst inaccuracies, we see \tool{} as playing a role in supporting more exploratory tasks where the potential damage of misunderstanding is limited, and in being consulted alongside up-to-date documentation for higher-stakes development tasks.
\section{Conclusion}

In this paper, we propose the notion of an instructive copilot,  a generative AI assistant that provides just-in-time explanations of its generations. We explore this idea in the setting of programming assistants, developing a tool called \tool{} that explains unfamiliar APIs in generated code. Our goals in designing \tool{} were to provide explanations that were anchored to expressions in generated code, lightweight, easy to invoke, easy to dismiss, and accessible anytime. In a usability study, \tool{} led to better comprehension of unfamiliar APIs in generated code versus a chat AI baseline. \tool{} also reduced task load and self-reported distraction. Programmers preferred \tool{} and saw \tool{}'s concise explanations as complementary to longer-form programming help like AI chat aids. Furthermore, our study revealed opportunities to improve explanations by making them expandable, adaptable, and configurable. Altogether, this work shows the value of lightweight, anchored AI support as a tool in the programming help-seeking toolkit.

\begin{acks}
We thank our colleagues from Penn HCI for their feedback on prototypes,  study designs, and paper drafts. We also thank Sorin Lerner for advice on how to implement overlays for VSCode.
\end{acks}

\bibliographystyle{ACM-Reference-Format}
\bibliography{references}
\balance{}
\appendix
\section{Timed comprehension assessment}
\label{apx:timed}

The comprehension assessments asked two kinds of questions:

\begin{itemize}
    \item \textbf{Identify the API}. The participant was told to imagine that they were trying to achieve a particular image manipulation goal and then shown the name of an API. They were asked whether the API could be used to achieve that goal. 8 questions were shown, 1 with the correct API from the programming task, and 7 with incorrect APIs either entirely made up or taken from the OpenCV documentation. 
    \item \textbf{Identify the purpose of a parameter}. The participant was shown the API signature without parameter names (e.g., ``\texttt{cv.Canny(\ding{182}, \ding{183}, \ding{184})}'').  They were directed to a specific parameter (e.g., ``\ding{182}''), shown a phrase (e.g., ``higher threshold'') and asked if the phrase described that parameter. 4 questions were shown for each parameter, with 1 correct description and 3 incorrect ones. We asked 12 such questions.
\end{itemize}

The test interface and procedure was designed so that response time measured only time to think and respond, to the extent possible. Responses were entered by pressing numeric pad keys ``1'' (for yes), ``2'' (for no), or ``3'' (for unsure), and then the Enter key to confirm. The programmer was trained in this system on sample questions before answering any of the questions we planned to analyze. They were told to answer questions as quickly as they were able. Screenshots of the assessment interface and a listing of all questions can be viewed in the supplemental material.

\section{Open-ended programming task}
\label{apx:game}

For the open-ended programming task, participants were asked to create a lightweight version of the classic Mario platformer game using Pygame~\cite{pygameor80:online}, a 2D game development library for Python. Participants were given no starter code. Rather, they were provided with a blank code Python script, a terminal from which they could run that script, and a folder containing graphics they could use in their game. Links to these graphics are listed in the supplemental material. Participants had access to \tool{}, EasyCode (the baseline chat AI), and the web. The task was designed to require programmers to see a significant number of unfamiliar API methods and Pygame-specific programming idioms. 78\% participants reported having no prior experience with Pygame, and 22\% reported being beginners. Most participants were unfamiliar with the domain of game development---72\% were not at all familiar, 19\% a little familiar, and 9\% somewhat familiar. Participants were not expected to finish, but rather to just make some progress. About half of participants got to a stage of development where a Mario sprite appeared on the screen and could be controlled with arrow keys.

\begin{figure*}[t!]%
    \centering
    \includegraphics[width=0.6\linewidth]{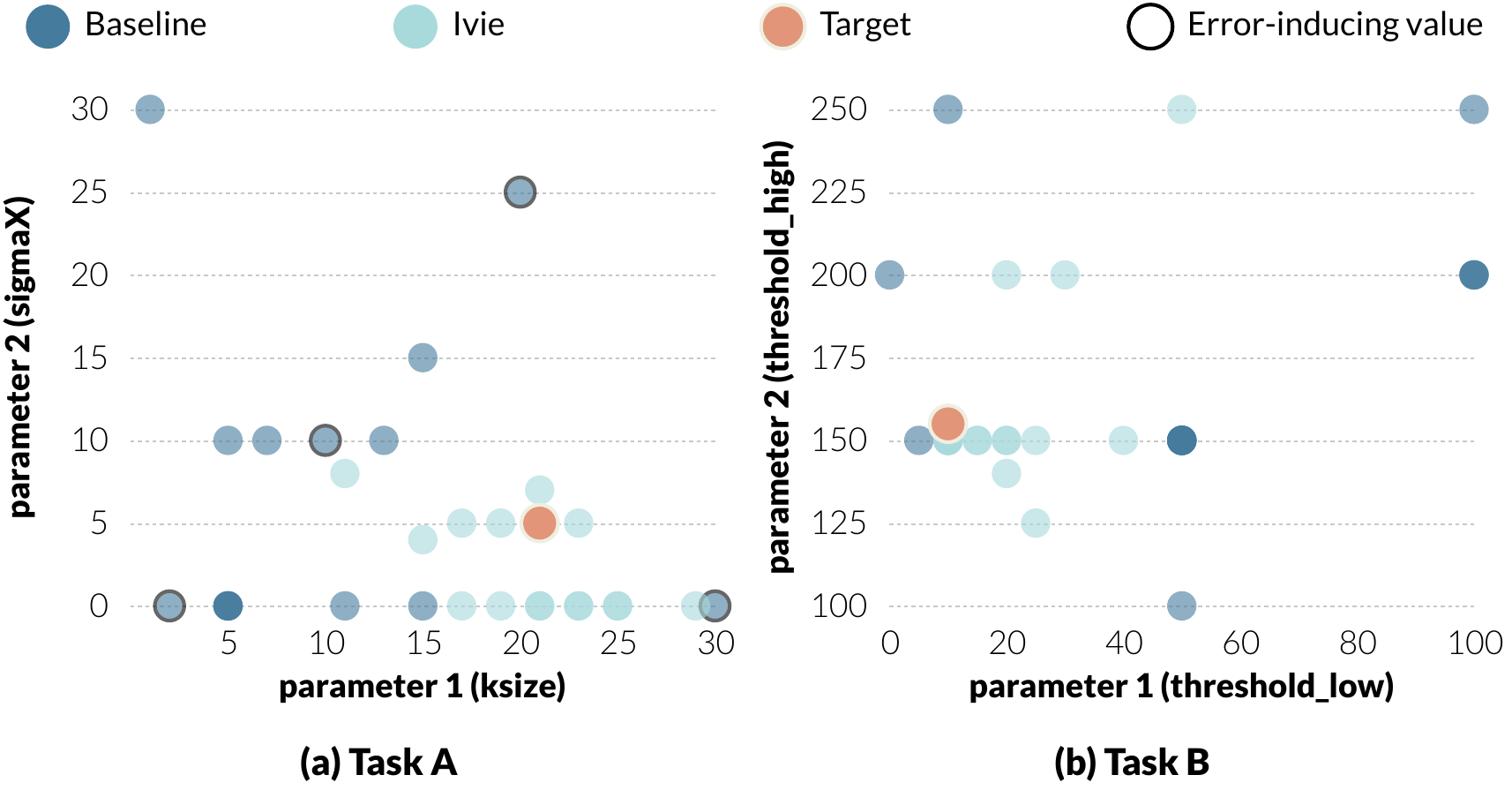}
    \vspace{-1ex}
    \caption{Convergence on parameter values. \textmd{Shown are plots conveying how far participants were from the reference solution when they ran out of time in each timed programming task. Each plot visualizes a parameter space for the key API function that the programmer need to use in a task. Each task required configuration of two parameters. All parameters are 1-dimensional numeric, except for ``\texttt{ksize}'' (``parameter 1'' in (a)), which was a tuple of two values, which were often set to be equal; we show only one of the values from the tuple. Red dots mark target configurations, which are (21, 5) for task A and (10, 155) for task B. Light blue dots represent values achieved when programmers used \tool{}, and dark blue plots represent values achieved in the baseline condition. Some points overlap: for instance, there are 5 overlapping baseline points in (a) at (5, 0); 8 overlapping baseline points in (b) at (50, 150) (this was frequently the default initial generated configuration); and 3 overlapping baseline points in (b) at (100, 200).}
    }%
    \label{fig:value}%
     \Description[]{2D scatter plots showing the proximity of arguments in participants’ code to the values in the reference solution. The plots show a rough trend where, when participants used Ivie instead of the baseline, their arguments tended to be closer to those in the reference solution.}
\end{figure*}

\section{Assessing Progress on Timed Programming Tasks}
\label{apdex:a}

A supplementary measure of task success was the extent to which participants' final code resembled a reference implementation. Each task required programmers to configure a set of parameters for an image processing API to control visual effects like blur or edge tracing to replicate a target image. In both the \tool{} and baseline condition, we collected the values of parameters in participants' code at the time they were cut off. For 2 of 4 parameters, participants were significantly closer to the target values when they used \tool{} versus using the baseline (see Figure~\ref{fig:value}). For the other 2 parameters, the differences were not significant. Significance was assessed by conducing an unpaired $t$-test of the L1 distance of parameter values to the target values for 1D parameters, and L2 distance for 2D parameters (e.g., the \ttt{ksize} tuple). Detailed results are as follows: For task A, there were two parameters. The first parameter, ``\texttt{ksize},'' had a target value of \texttt{(21, 21)}. The L2 distance of programmers' final parameters to this target value was 1.41 ($\sigma = 11.8$) in the \tool{} condition versus 14.8 ($\sigma = 16.3$) in the baseline condition; this difference was statistically significant ($p = 0.004$). No significant difference was seen for the distance to the target value for the second parameter ``\texttt{sigmaX}'' between \tool{} ($\mu = 2.56$, $\sigma = 2.89$) and the baseline ($\mu = 7.81$, $\sigma = 29.84$; $p = 0.19$). For task B, programmers were significantly closer to the target value of the ``\ttt{threshold\_low}'' parameter in the \tool{} condition ($\mu = 10.62$, $\sigma = 11.16$) than in the baseline condition ($\mu = 50.31$, $\sigma = 46.38$; $p = 0.003$), though not for the for the ``\ttt{threshold\_high}'' parameter ({\tool}: $\mu = 8.44$, $\sigma = 39.79$ vs. baseline: $\mu = 20.0$, $\sigma = 46.77$; $p = 0.472$).

\end{document}